\documentclass[%
reprint,
 amsmath,amssymb,
 aps,
]{revtex4-2}

\usepackage{graphicx}% Include figure files
\usepackage{dcolumn}% Align table columns on decimal point
\usepackage{bm}% bold math
\usepackage{amsmath,MnSymbol}

\usepackage{xcolor}
\usepackage{bbding}

\begin{document}

\preprint{APS/123-QED}

\title{Kinematic and rheological equivalence of steady shearing and planar extensional flows }% Force line breaks with \\

\author{Nicholas King}
\affiliation{Department of Chemical Engineering, Massachusetts Institute of Technology,
Cambridge, Massachusetts 02139, USA}%

\author{Gareth H. McKinley}
\affiliation{Hatsopoulos Microfluids Laboratory, Department of Mechanical Engineering, Massachusetts Institute of Technology, Cambridge, Massachusetts 02139, USA}%

\date{\today}% It is always \today, today,

\begin{abstract}
Steady shearing and planar extension are commonly viewed as two distinct types of flow field, especially in the context of probing the rheology of complex fluids. By leveraging the kinematic equivalence between the two flows, we derive an effective extension rate experienced by a material element which removes the rotational component of the shearing flow. This enables reconstruction of the steady planar extensional viscosity of an unknown fluid using only material functions measured in a steady shearing flow, revealing a deep rheological equivalence between the two deformation histories. We demonstrate this equivalency through phenomenological and microscopically motivated frame-invariant constitutive models as well as using experimental data obtained with a viscoelastic polymer solution and a polymer melt.
\end{abstract}

\maketitle

%\tableofcontents

The nonlinear rheology of complex fluids is typically characterized in homogeneous flow fields such as simple shear and pure extension. These two flows probe different aspects of the microstructure through different stretch histories, and in general, one cannot be predicted from the other \cite{petrie_one_2006,larson_modeling_2015}. Although shear flows are routinely employed in a torsional rheometer to obtain material functions such as the shear viscosity and the first normal stress coefficient, steady extensional flows have been more difficult to achieve \cite{petrie_extensional_2006}. Nevertheless, extensional flows are essential for understanding many polymer processing operations \cite{huang_when_2022}, as well as filament breakup \cite{mckinley_filament-stretching_2002}, fracture \cite{lima_unexpected_2026} and jetting phenomena \cite{keshavarz_studying_2015}. Furthermore, in flows with mixed kinematics, the extensional component will often dominate at high strain rates \cite{petrie_elongational_1979}. 

% Of the different types of extensional flows, i
It is commonly assumed that planar elongation probes nonlinear material responses that are independent of those probed by shearing deformations \cite{laun_transient_1989}. However, generating true planar extensional flows has posed a major challenge experimentally, especially compared to implementing uniaxial or biaxial extensional deformations. Various devices have been developed for generating uniaxial elongation, though these deformations are typically transient and inhomogeneous. Examples include the use of converging flows \cite{laun_transient_1989} or the rotating clamp Meissner apparatus \cite{hachmann_rheometer_2003} for viscous polymer melts, and filament stretching configurations \cite{mckinley_filament-stretching_2002} such as Dripping onto Substrate \cite{dinic_extensional_2015}, and jetting devices such as ROJER \cite{keshavarz_studying_2015} for less viscous materials. The only bespoke devices able to generate planar extensional flow for dilute polymer solutions are the OSCER microfluidic device \cite{haward_optimized_2012} and a modified filament stretching device \cite{nguyen_response_2015}. The difficulties in generating planar extensional flow, especially at sufficiently high extension rates to achieve large microstructural deformations, have led to few published studies.

More broadly, flow fields can be classified kinematically using invariants of frame-invariant tensors such as the strain rate $\pmb{\dot{\gamma}}$ and the Finger strain $\pmb{B}$. Because of fluid incompressibility, the first invariant of $\pmb{\dot{\gamma}}$ is zero, and the third invariant of $\pmb{B}$ is always unity. The remaining two invariants for each tensor thus form two separate maps of all physically realizable flows. Such maps were first put forth by Sawyers \cite{sawyers_possible_1977} in terms of the Finger strain tensor, and have been popularized by Chong, Perry and Cantwell \cite{chong_general_1990} for classifying turbulent flow fields in terms of the strain rate tensor. Both representations are featured in the classic textbook by Bird \textit{et al.} \cite{bird_dynamics_1987} and are summarized in the SI. These maps are bounded by uniaxial and biaxial extensional flows, but interestingly, simple shear flow and planar extension reside on the same invariant lines and are thus kinematically indistinguishable. This raises a more fundamental question about whether a mapping exists between steady shear material functions and the steady planar extensional viscosity. In this Letter, we show that for any materially objective constitutive equation, 
%which admits steady state solutions in simple shear, 
the steady planar extensional viscosity can be completely determined from the steady shear viscosity and the first normal stress difference. This relationship reveals that the two flows, although ostensibly distinct, are kinematically equivalent. Our key development is to systematically remove the contribution of rotation from the shearing flow, by defining an ``effective extension rate" to establish this equivalence. This establishes an accessible route to evaluate steady planar extensional behavior from steady shear measurements alone, and opens avenues for understanding the physics inherent in a complex fluid subjected to two flows previously thought to be distinct.

In a recent study, Hillebrand \textit{et al.} \cite{hillebrand_flat_2025} have shown that suspensions of elastic discs and solutions of long flexible polymers, described respectively by the Oldroyd-A and Oldroyd-B models, produce identical velocity fields in any planar 2D flow. Our finding builds on this development, since both constitutive models generate identical shear stress and first normal stress differences in shear flow, and we show more broadly that the rheological material functions defined in planar extensional flows are kinematically connected to those in steady shear flow, once the rate of rotation has been accounted for.

By symmetry arguments, the extra stress tensor $\pmb{\sigma}$ in any planar flow (i.e. including simple shear and planar extensional flow) has the following form:
\begin{equation}
    \pmb{\sigma} = 
    \begin{pmatrix}
        \sigma_{11} & \sigma_{12} & 0 \\
        \sigma_{12} & \sigma_{22} & 0 \\
        0 & 0 & \sigma_{33}
    \end{pmatrix}
\end{equation}

Meaningful material functions are always defined in terms of the system's  response (i.e. the output) divided by a measure of the imposed deformation history (input) \cite{ad_hoc_committee_on_official_nomenclature_and_symbols_official_2013}. For steady shear, we define 1 as the flow direction and 2 as the velocity gradient direction (Fig. 1(a)). The only non-zero component of the velocity field is $v_1 = \dot{\gamma} x_2$. Hence, the shear viscosity is defined as $\eta(\dot{\gamma}) \equiv \sigma_{12} / \dot{\gamma}$, and rheologists also measure the first normal stress difference $N_1 = \sigma_{11} - \sigma_{22}$ and the second normal stress difference $N_2 = \sigma_{22} - \sigma_{33}$. For steady homogeneous planar extensional flow, we define 1 as the extension direction and 2 as the compression direction (Fig. 1(b)). The velocity field has two non-zero components $v_1 = \dot{\epsilon} x_1$ and $v_2 = -\dot{\epsilon} x_2$. The steady planar extensional viscosity is defined by $\eta_{P1} (\dot{\epsilon}) \equiv (\sigma_{11} - \sigma_{22}) / \dot{\epsilon}$. For completeness, we also define a second planar extensional viscosity $\eta_{P2} (\dot{\epsilon}) \equiv (\sigma_{33} - \sigma_{22}) / \dot{\epsilon}$. 
% Note that from symmetry arguments based on reversing the flow direction (such that $\dot{\gamma} \rightarrow - \dot{\gamma}$ or $\dot{\epsilon} \rightarrow - \dot{\epsilon}$), we expect that the normal stress difference $N_1$ in steady shear is an even function of the applied shear rate, $(\sigma_{11} - \sigma_{22}) \sim \dot{\gamma}^2$, whereas the corresponding quantity in planar extensional flow is an odd function of the imposed extension rate, i.e. $(\sigma_{11} - \sigma_{22}) \sim \dot{\epsilon}$ \cite{bird_dynamics_1987}. 

The extra stress tensor for any planar flow can also be expressed in terms of three principal stresses $\sigma_1$, $\sigma_2$ and $\sigma_3$ when rotated into its principal axes:
\begin{equation} \label{rotatedstress}
    \pmb{\sigma}' = 
    \begin{pmatrix}
        \sigma_{1} & 0 & 0 \\
        0 & \sigma_{2} & 0 \\
        0 & 0 & \sigma_{3}
    \end{pmatrix}
%    \begin{pmatrix}
%        \frac{1}{2} \left( \sigma_{11} + \sigma_{22} + \sqrt{(\sigma_{11} - \sigma_{22})^2 + 4\sigma_{12}^2} \right) & 0 & 0 \\
%        0 & \frac{1}{2} \left( \sigma_{11} + \sigma_{22} - \sqrt{(\sigma_{11} - \sigma_{22})^2 + 4\sigma_{12}^2} \right) & 0 \\
%        0 & 0 & \sigma_{33}
%    \end{pmatrix}
\end{equation}

In planar extensional flow, the stress tensor is already diagonal since $\sigma_{12} = 0$, hence $\sigma_1 = \sigma_{11}$, $\sigma_2 = \sigma_{22}$ and $\sigma_3 = \sigma_{33}$. It is frequently described by rheologists as a `shear-free' flow \cite{bird_dynamics_1987}, but is also referred to as `pure shear' by solid mechanicians \cite{anand_continuum_2020}. In simple shear flow, the eigenvalues of $\pmb{\sigma}$ are \cite{lodge_elastic_1964}:
\begin{subequations}
\begin{equation}
    \sigma_1 = \frac{1}{2} \left( \sigma_{11} + \sigma_{22} + \sqrt{(\sigma_{11} - \sigma_{22})^2 + 4\sigma_{12}^2} \right) 
\end{equation}
\begin{equation}
    \sigma_2 = \frac{1}{2} \left( \sigma_{11} + \sigma_{22} - \sqrt{(\sigma_{11} - \sigma_{22})^2 + 4\sigma_{12}^2} \right) 
\end{equation}
\begin{equation}
    \sigma_3 = \sigma_{33}
\end{equation}
\end{subequations}

When shearing deformations are expressed in this coordinate frame, we can define a principal stress difference $\Delta \sigma \equiv (\sigma_1 - \sigma_2) = \sqrt{N_1^2 + 4 \sigma_{12}^2}$ (see Fig. 1(c)). As the shear flow strength increases, the microstructure becomes increasingly aligned with the 1 direction, and this can be quantified by an orientation angle $\chi = \tfrac{1}{2} \tan^{-1} (2\sigma_{12} / N_1)$ that decreases from 45$^\circ$ towards 0$^\circ$ \cite{lodge_elastic_1964}. The question is thus: can we define a composite viscosity function that maps the material responses determined in shear and is consistent with the true planar extensional viscosity $\eta_{P1} = (\sigma_{11} - \sigma_{22}) / \dot{\epsilon}$ defined in Fig. 1? 

\begin{figure} [h]
    \centering
    \includegraphics[width=0.5\textwidth]{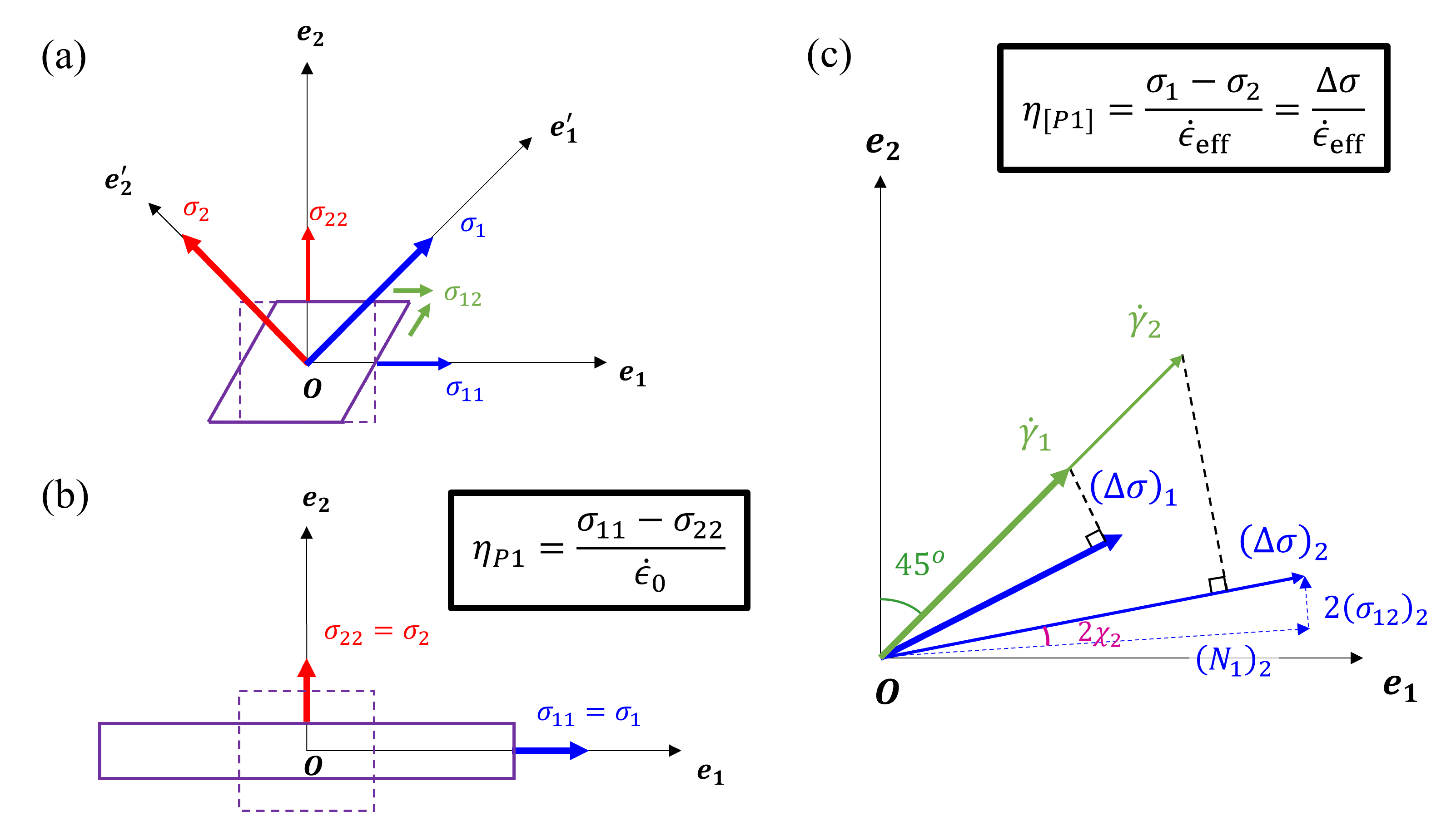}
    \caption{(a) The different stresses $\sigma_{11}$, $\sigma_{22}$ and $\sigma_{12}$ acting on a fluid element in steady shear flow. The principal stresses $\sigma_1 (t)$ and $\sigma_2 (t)$ and the corresponding eigenvectors ${\bf e_1'}$ and ${\bf e_2'}$ are shown at an arbitrary time $t$ for a given flow strength $\dot\gamma$. (b) The principal stresses $\sigma_1$ and $\sigma_2$ at time $t$ in a planar extensional flow. The orientation of the eigenvectors do not evolve in time. (c) A schematic depicting the projection of the imposed strain rate onto the principal axes of stress in shear flow for two different flow strengths $\dot\gamma_1, \dot\gamma_2$. The principal stress difference at each instance is $\Delta \sigma = \sigma_1 - \sigma_2 = \sqrt{N_1^2 + 4 \sigma_{12}^2}$.}
\end{figure}

To answer this question, we define an \textit{effective extension rate} $\dot{\epsilon}_{\mathrm{eff}}$ by projecting the imposed deformation rate field onto the evolving principal axes of the resulting stress tensor. For general deformations, the symmetric strain rate tensor is defined as $\pmb{\dot{\gamma}}= \nabla {\bf v} + \nabla {\bf v}^T$, in which $\nabla {\bf v}$ is the velocity gradient tensor. In planar extensional flow, this tensor is diagonal and can be expressed compactly as $\pmb{\dot{\gamma}} = 2 \dot{\epsilon} {\bf e_1 e_1} - 2\dot{\epsilon} {\bf e_2 e_2}$. Therefore, the eigenvectors of the strain rate tensor and the stress tensor (cf. Eq. \ref{rotatedstress}) are always coincident and fixed with time. On the other hand, the strain rate tensor in simple shear is $\pmb{\dot{\gamma}} = \dot{\gamma} {\bf e_1 e_2} + \dot{\gamma} {\bf e_2 e_1}$. As we show schematically in Fig. 1(c), the eigenvectors of the strain rate tensor are always fixed in time (even in a transient shear experiment) and rotated 45$^\circ$ anticlockwise from the flow axis ${\bf e_1}$. However, the eigenvectors of the stress tensor change their orientation at a rate that is dependent on the evolution of the stress field given by the material's constitutive response. The full expressions for the time-dependent eigenvectors of the stress tensor are provided in the SI. Since the principal axes of the strain rate tensor and the principal axes of the stress tensor are not coincident, the proper way to define the effective extension rate that is aligned in the direction of the stress is to find the projection of the imposed deformation rate onto the direction of the principal axes of the stress tensor, as we sketch in Fig. 1(c).

We define the normalized eigenvectors of the stress tensor as $\bf {e}_1'$, $\bf {e}_2'$ and $\bf {e}_3'$. 
%, which form an orthonormal set of base vectors. 
By rotating the strain rate tensor into this coordinate frame and extracting the 11 component of the resultant tensor, the ``effective extension rate" can be shown to be (see details in the SI):
\begin{equation} \label{eff}
    \dot{\epsilon}_\mathrm{eff} = \frac{1}{2} \pmb{e}_1' \cdot \pmb{\dot{\gamma}} \cdot \pmb{e}_1' = \frac{\dot{\gamma} \sigma_{12}}{\Delta \sigma}
\end{equation}

This is the main finding of this Letter. The factor of $1/2$ arises because we constrain the ``flow strength", measured by the second invariant of the strain rate tensor, to be equivalent for both simple shear flow and planar extension. In simple shear flow, $II_{\pmb{\dot{\gamma}}} = \text{tr} (\pmb{\dot{\gamma}}^2) =  2 \dot{\gamma}^2$, while in a general planar extension, $II_{\pmb{\dot{\gamma}}} = 8 \dot{\epsilon}^2$. Equating the two flow strengths, for internal consistency we have $\dot{\epsilon}_\mathrm{eff} = \dot{\gamma}/2$. %For internal consistency, this accounts for the factor of 1/2 even after projecting the strain rate tensor onto the eigenvectors of the stress tensor.

This effective extension rate $\dot{\epsilon}_\mathrm{eff}$ now isolates the stretching component of the shear flow projected onto a deforming material element at time $t$ by removing the vorticity inherent to a simple shearing deformation. This then leads to a proper definition of an equivalent composite planar extensional viscosity:
\begin{equation} \label{etap1}
    \eta_{[P1]} \equiv \frac{(\sigma_1 - \sigma_2)}{\dot{\epsilon}_\mathrm{eff}} 
    %= \frac{\Delta \sigma}{\dot{\epsilon}_{eff}}
    = \frac{N_1^2 + 4 \sigma_{12}^2}{\dot{\gamma} \sigma_{12}} = 4 \eta(\dot\gamma)\left[1+\left( \frac{N_1}{2\sigma_{12}} \right)^2\right]
\end{equation}
where we have substituted $\eta(\dot\gamma)\equiv \sigma_{12}/ \dot\gamma$ in the last equality. The ratio $\tfrac{N_1}{2\sigma_{12}}$ represents the ``recoverable elastic strain" in the fluid and is the original basis \cite{white_dynamics_1964} for the definition of the Weissenberg number \cite{poole_deborah_2012}. We show that when this new composite function is plotted against the effective extension rate, we can completely reconstruct the \textit{a priori} unknown steady planar extensional viscosity curve, using only steady shear flow data.

An expression for the second planar extensional viscosity can also be obtained by dividing the appropriate principal stress difference by the effective extension rate:
\begin{equation}
    \eta_{[P2]} \equiv \frac{(\sigma_3 - \sigma_2)}{\dot{\epsilon}_\mathrm{eff}} = \frac{(\Delta \sigma) (\Delta \sigma -N_1 - 2N_2)}{2 \dot{\gamma} \sigma_{12}}
\end{equation}

However, Petrie \cite{petrie_asymptotic_1990} has shown that $\eta_{P2}$ becomes vanishingly small compared to $\eta_{P1}$ in planar extension for most rheological models. Furthermore, it is more difficult to measure $N_2$ in a rheometer \cite{maklad_review_2021} especially since its magnitude is often much smaller than $N_1$. In view of these limitations, the rest of this Letter focuses on the dominant planar extensional viscosity $\eta_{[P1]}$. Further results on $\eta_{[P2]}$ are provided in the SI.

Inelastic fluids exhibit no normal stress differences, \( N_1 = N_2 = 0\) and no recoverable elastic strain. In the Newtonian limit, $\sigma_{12} = \eta_0 \dot{\gamma}$ in shear flow, and the planar extensional viscosity $\eta_{[P1]}$ predicted from Eq. \ref{etap1} is four times the shear viscosity $\eta_0$, in agreement with the well-known Trouton ratio $Tr = \eta_{P1} / \eta_0 = 4$ in planar extensional flow. Similarly, due to its inelasticity ($N_1 \rightarrow 0$), a generalized Newtonian fluid will always have a planar extensional viscosity that is four times the shear viscosity, even if the fluid shear thins. Numerical simulations on a suspension of solid spherical particles, which flows like an inelastic fluid, in planar extension \cite{cheal_rheology_2018} have also shown that the Trouton ratio remains at four at large extension rates. 

\begin{figure} [h]
    \centering
    \includegraphics[width=0.48\textwidth]{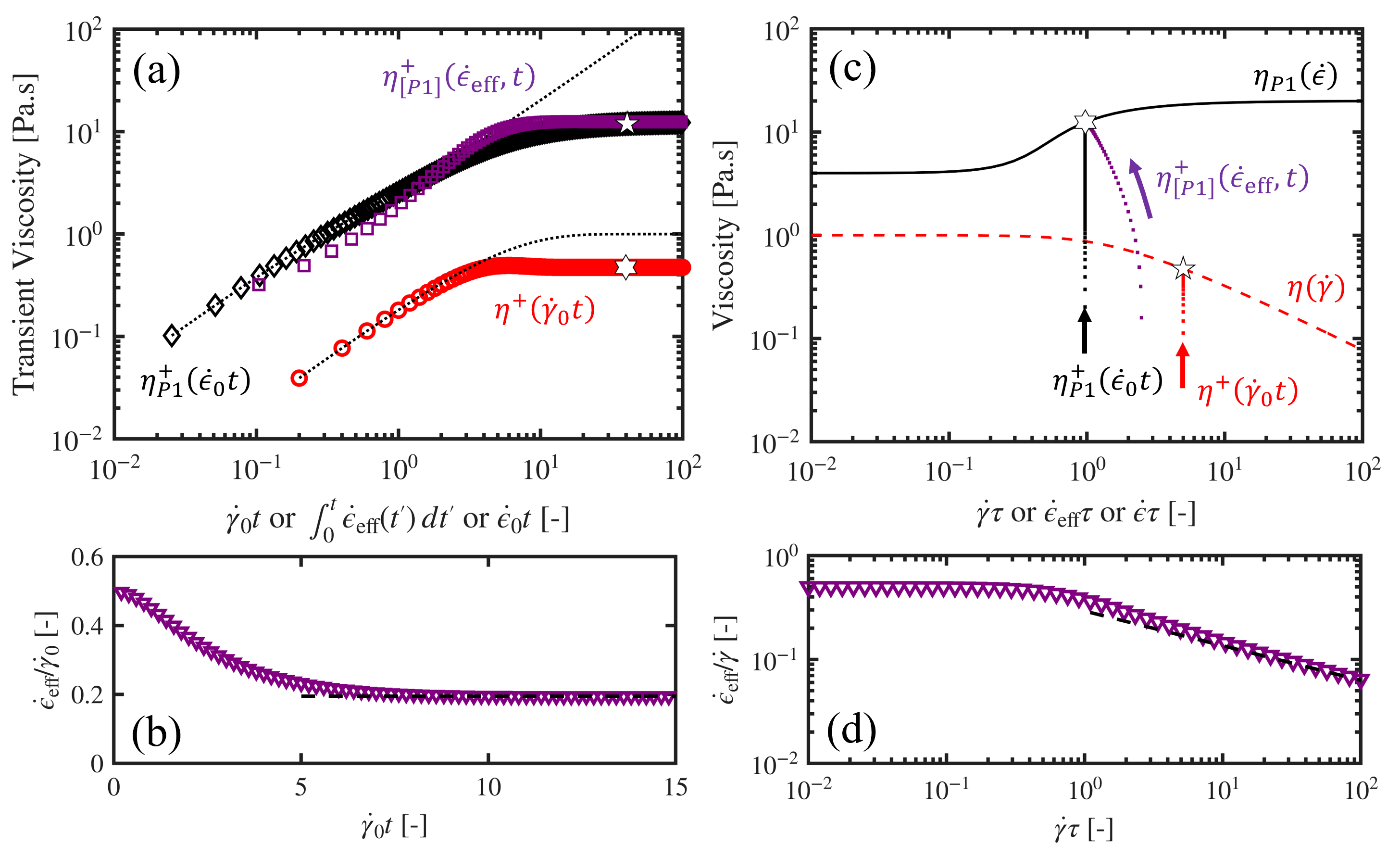}
    \caption{(a) Transient shear and planar extensional viscosity for the linear PTT model ($\eta_0 = 1$ Pa.s, $\tau = 1$ s, $\varepsilon = 0.1$) as a function of accumulated strain. A step shear rate $\dot{\gamma}_0 = 5$ s$^{-1}$ or an extension rate $\dot{\epsilon}_0 = 0.97$ s$^{-1}$ was applied. The predictions of the UCM model are shown as dotted lines. The steady state values of these curves (star symbols) correspond to a single point on the graph of the steady viscosities, shown in (c). The variation of the effective extension rate, normalized by the applied strain rate, are shown against (b) time, and (d) strain rate.}
\end{figure}

With the insight of Eq. \ref{etap1}, we are able to replicate the steady state planar extensional viscosity curves for a wide range of constitutive models. We first show that this composite material function is valid for a phenomenological nonlinear model based on polymer network formation and destruction, the Phan-Thien Tanner (PTT) model \cite{phan-thien_new_1977}. Specifically, the linear PTT model has similarities to the widely used FENE-P model for polymer solutions when the flows are steady \cite{davoodi_similarities_2022}. It has the equation:
\begin{equation}
    \tau \overset{\kern0.25em\smalltriangledown}{\pmb{\sigma}} + \left(1 + \frac{\varepsilon \tau}{\eta_0}\text{tr}(\pmb{\sigma}) \right) \pmb{\sigma} = \eta_0 \dot{\pmb{\gamma}}
\end{equation}
where the operator ($\overset{\kern0.25em\smalltriangledown}{.}$) indicates the upper convected derivative $\overset{\kern0.25em\smalltriangledown}{\pmb{\sigma}} \equiv \frac{D \pmb{\sigma}}{D t} - (\nabla \pmb{v})^T \cdot \pmb{\sigma} - \pmb{\sigma} \cdot \nabla \pmb{v}$, and $\tau$ is the relaxation time of the fluid. The model nonlinearity is controlled by a parameter $\varepsilon$, which is a measure of the enhanced rate of rupture events in the transient entangled network, and the model simplifies to the quasilinear Upper Convected Maxwell (UCM) model when $\varepsilon = 0$. To solve for the stress tensor $\pmb{\sigma}$, we consider the start-up of steady shear $\dot{\gamma}(t) = \dot{\gamma}_0 \mathcal{H}(t)$ and planar extensional flow $\dot{\epsilon}(t) = \dot{\epsilon}_0 \mathcal{H}(t)$, where $\mathcal{H}(t)$ is the Heaviside step function. We define an accumulated Hencky strain $\int_0^t \dot{\epsilon}(t') \, dt'$ for extensional flow, and an accumulated strain for shear flow $\int_0^t \dot{\gamma} (t') \, dt' = \dot{\gamma}_0 t$. As we show in Fig. 2(a), the shear stress growth coefficient \cite{ad_hoc_committee_on_official_nomenclature_and_symbols_official_2013} $\eta^+ = \sigma_{12}(t) / \dot{\gamma}_0$ evolves nonlinearly with shear strain and saturates. In extensional flow the stress difference is larger, but $\eta_{P1}^+$ also saturates at large Hencky strains. Our new composite material function $\eta_{[P1]}^+$ closely follows this material response and, importantly, saturates at exactly the same value. The dimensionless ratio of the extension rate $\dot{\epsilon}_\mathrm{eff}(t) / \dot{\gamma}_0$ (Fig. 2(b))  decreases from an initial value of 1/2 and plateaus to a steady value at large strains. The predictions from the UCM model, which provides the universal asymptotic limits at low strain rates, are shown as dotted lines, emphasizing the transition from linear to nonlinear fluid response at high shear rates ($\dot{\gamma} \tau \gg 1$) and large strains. The UCM model does not predict a steady extensional viscosity at dimensionless extension rates beyond $\tau \dot{\epsilon}_0 > \tfrac{1}{2}$.  
%When evaluating the composite material function, the eventual effective extension rate that will be achieved cannot be predicted \textit{a priori} because it depends on the evolution of the shear stress and $N_1$ in the start up of steady shear, which are set by the response of the specific constitutive equation. 
% of the effective extension rate realized to the imposed shear rate

We represent the steady state responses of this dynamical system over a wide range of deformation rates in Fig. 2(c). We show that the trajectories of the three transient stress growth coefficients each evolve towards the corresponding steady state attractors. Importantly, our new composite material function $\eta_{[P1]}$ (determined from transient shear properties alone) converges to the same value as $\eta_{P1}$ (obtained from homogeneous planar extension). This is because the effective extension rate begins at $\dot{\gamma}/2$ at low strain rates, and later decreases (in a model-dependent way) because $\dot{\epsilon}_\mathrm{eff} / \dot{\gamma} \rightarrow \sigma_{12} / N_1$ at high strain rates (Eq. \ref{eff}). Asymptotic analysis of the linear PTT model at large strain rates \cite{shogin_full_2021} gives $\sigma_{12} \sim \eta_0 (2 \varepsilon \tau^2)^{-1/3} \dot{\gamma}^{1/3}$ and $N_1 \sim \eta_0 (\varepsilon^{2} \tau / 2)^{-1/3} \dot{\gamma}^{2/3}$, hence as shown by the dashed line in Fig. 2(d), $\dot{\epsilon}_\mathrm{eff} / \dot{\gamma} \sim (4/\varepsilon)^{-1/3} (\tau \dot{\gamma})^{-1/3}$.

These conclusions are further supported by numerical calculations (Fig. 3(a)). The planar extensional viscosity $\eta_{P1} (\dot{\epsilon})$ increases due to the coil-stretch transition \cite{de_gennes_coil-stretch_1974} at an extension rate corresponding to a critical Weissenberg number $Wi_c = \tau \dot{\epsilon}_c = \tfrac{1}{2}$. The material response softens at higher extension rates, causing $\eta_{P1}$ to saturate smoothly. 
% This indicates that we can effectively generate the steady planar extensional viscosity curve using only steady shear data. 

Our analysis also applies to microstructurally motivated polymer models. We illustrate this by considering a state-of-the-art continuum model for entangled polymer melts, the Rolie-Poly model \cite{likhtman_simple_2003}. In terms of the polymer configuration tensor $\pmb{c}$ and the identity tensor $\pmb{I}$, the Rolie-Poly model is written as:
\begin{equation}
    \overset{\kern0.25em\smalltriangledown}{\pmb{c}} = - \frac{1}{\tau_d} (\pmb{c} - \pmb{I}) - \frac{2 k_s (\lambda)}{\tau_R} \left(1 - \frac{1}{\lambda}\right) \left(\pmb{c} + \beta \lambda^{2\delta} (\pmb{c} - \pmb{I}) \right)
\end{equation}
where the disengagement time is $\tau_d$, and the chain stretch $\lambda = \sqrt{\text{tr}~\pmb{c} / 3}$ evolves separately with the Rouse time $\tau_R$. The number of entanglements per polymer chain $Z$ is often used to describe the degree of entanglements in the melt, and controls the ratio $\tau_d / \tau_R$ \cite{likhtman_quantitative_2002,du_capillarity-driven_2025}. Two scalar model parameters $\beta$ and $\delta$ modulate the effectiveness of convective constraint release and the chain sensitivity to stretching \cite{holroyd_analytic_2017}. The finite extensibility of the polymer chains is incorporated using the Cohen-Padé approximation to the inverse Langevin function for the spring law \cite{kabanemi_nonequilibrium_2009}, making the spring constant $k_s(\lambda)$ a function of the maximum extensibility ($L_{max}$) of the polymer chain (refer to SI for details). The tensorial stress is related to the polymer configuration tensor $\pmb{c}$, the nonlinear spring function $k_s(\lambda)$ and the plateau modulus $G$:
\begin{equation}
    \pmb{\sigma} = G k_s(\lambda) (\pmb{c} - \pmb{I})
\end{equation}
The entangled fluid described by the Rolie-Poly model exhibits shear thinning of the shear viscosity beyond $Wi_d \equiv \dot{\gamma} \tau_d \gtrsim 1$ (Fig. 3(b)). Once again, our proposed composite material function $\eta_{[P1]}$ accurately replicates the entire planar extensional viscosity curve. At low $Wi_d$, $N_1$ increases quadratically, but when $N_1 \gtrsim \sigma_{12}$, nonlinear elastic effects grow substantially and $\eta_{[P1]}$ evolves non-monotonically. Physically, the initial decrease in $\eta_{P1}$ is due to the deformation-induced orientation of the chains within their tubes, followed by a coil-stretch transition and a sharp increase in the tensile stress difference at a critical Weissenberg number based on the Rouse time $\dot{\epsilon}_c \tau_R \sim O(1)$ (or equivalently, $Wi_c \sim \tau_d / \tau_R \sim 3Z$). Finally, $\eta_{P1}$ saturates due to finite extensibility effects and this is also faithfully captured by $\eta_{[P1]}$.
% It is nontrivial that $\eta_{[P1]}$ saturates at high $Wi_d$ given how $\sigma_{12}$, $N_1$ and $\dot{\epsilon}_\mathrm{eff}$ are all evolving nonlinearly. 

\begin{figure} [h]
    \centering
    \includegraphics[width=0.48\textwidth]{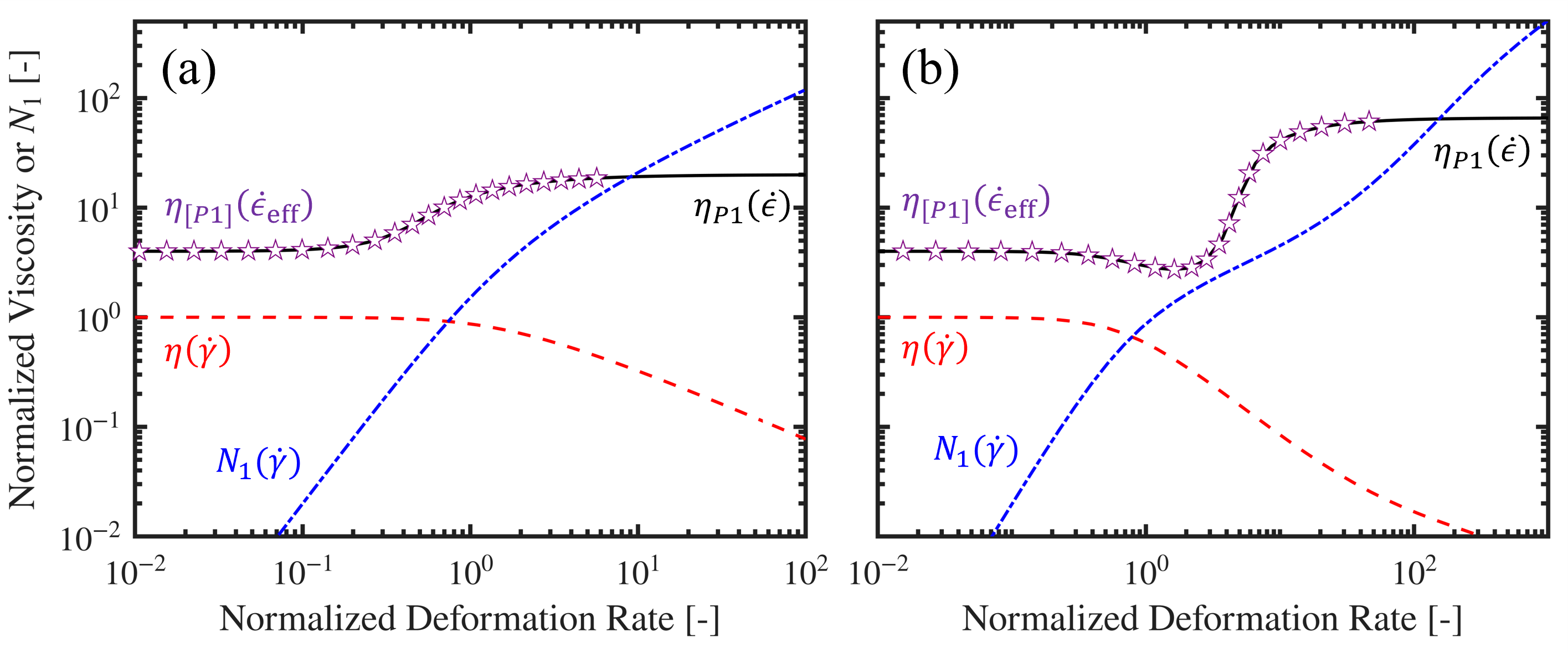}
    \caption{Steady planar extensional viscosity curves recreated using the steady shear $\eta(\dot{\gamma})$ and $N_1 (\dot{\gamma})$ for (a) the linear PTT model ($\varepsilon = 0.1$, $G=\eta_0/\tau$) and (b) the Rolie-Poly model ($\tau_d / \tau_R = 5$ s, $\beta = 0.1$, $\delta = -0.5$, $L_{max} = 10$). The lines are the viscosity $\eta(\dot{\gamma})$ ($\textcolor{red}{---}$), $N_1(\dot{\gamma})$ ($\textcolor{blue}{\cdot\!-\!\cdot}$) and the steady planar extensional viscosity $\eta_{P1}(\dot{\epsilon})$ (---). The symbols ($\textcolor{purple}{\text{\FiveStarOpen}}$) are the new composite material function $\eta_{[P1]}(\dot{\epsilon}_\mathrm{eff})$ (Eq. \ref{etap1}). All viscosities are normalized by $\eta_0$ and $N_1$ is normalized by the shear modulus $G$. The deformation rates for the linear PTT and for the Rolie-Poly model are normalized by $\tau$ and $\tau_d$ respectively.}
\end{figure}

In each of the examples above, simply observing either the flow curve of $\eta (\dot{\gamma})$ or the evolution of $N_1 (\dot{\gamma})$ with $\dot{\gamma}$ does not clearly reveal the existence of this stretching phenomenon because of the increasing angle between the eigenvectors of the imposed deformation rate and the resulting stress difference. Evaluating the new composite material function (Eq. \ref{etap1}) in terms of the measurable quantities in steady shear clearly reveals the physics of the coil-stretch transition, at no additional experimental cost. We show in the SI that $\eta_{[P1]}$ also fully captures the planar extensional flow response of other commonly used constitutive models such as the UCM, Giesekus, and exponential PTT models. We also show two representative examples using experimental data for a polymer melt and a semidilute polymer solution in the End Matter.

%Beyond this critical strain rate, elasto-inertial instabilities cause the sample to be ejected from the gap and measurements of the viscometric properties are no longer possible.

% It has recently been shown that fitting nonlinear data conjointly with linear data provides a more robust/credible fit \cite{singh_simultaneous_2022}. 
% $\dot{\gamma}_c \approx 80 s^{-1}$

%Nevertheless, since this prediction is grounded in the kinematic equivalence between steady shear and planar extensional flows, we expect it to hold. 

We have shown in this Letter that the steady planar extensional viscosity can be reconstructed purely from the rheological information contained in the steady shear viscosity and the first normal stress difference. This result follows from the congruity of the kinematics of the two flows (quantified in terms of both $II_{\pmb{\dot{\gamma}}}$ and $II_{\pmb{B}}$) and is independent of the material constitutive equation. We have illustrated this finding using classical nonlinear continuum models as well as a molecularly-informed reptation model. We have also demonstrated it using experimental data for a representative, weakly entangled, polymer solution which cannot be characterized in planar extension using other techniques. The dynamical response of a viscoelastic material in steady shear is entirely equivalent to that in planar extension when expressed in a suitable reference frame, but the principal stress difference is distributed between two material responses $\sigma_{12}$ and $N_1$. This is a step towards understanding planar extensional flows, and prompts a deeper re-evaluation of other previously published viscometric data in steady shear flows for which both $\sigma_{12}$ and $N_1$ have been carefully measured. 

Our new analysis augments the well-known Cox-Merz rule \cite{cox_correlation_1958} -- which states that for viscoelastic fluids, the nonlinear shear viscosity curve can be approximated by the magnitude of the complex viscosity obtained from linear viscoelasticity measurements as a function of angular frequency. The analogous concept here is that the nonlinear material response in one type of deformation (readily achievable in a torsional rheometer) enables prediction of the corresponding behavior in a very different (shear-free) flow. This stands alongside other empirical `rules' for complex fluids that relate nonlinear to linear viscoelastic behavior, such as Laun's rule \cite{laun_prediction_1986,das_launs_2024} and the Lodge-Meissner relation \cite{lodge_use_1972}. Moving forward, carefully resolved measurements of $\sigma_{12}(\dot{\gamma})$ and $N_1(\dot{\gamma})$ performed in steady shear flow over a range of shear rates can be used to construct steady planar extensional viscosity curves and provide greater insight into physical mechanisms underpinning strain-rate stiffening or softening, such as coil-stretch transitions and chain alignment, without incurring the additional costs and difficulties of performing more complicated planar extensional experiments. 
%Note that $\dot{\epsilon}_\mathrm{eff} (t)$ decreases as viscoelastic stresses grow and this needs to be measured (Fig. 1). 

\begin{acknowledgments}
We acknowledge Lubrizol Corporation for financial support and for providing the PIB fluid, as well as the Non-Newtonian Fluids group at MIT for the highly enjoyable reading group on flow kinematics in Fall 2025. N.K. also acknowledges conference grant support from the National University of Singapore Overseas Graduate Scholarship (NUS-OGS).
\end{acknowledgments}

\bibliography{effectiveextensionrate}% Produces the bibliography via BibTeX.

\clearpage

\appendix*
\renewcommand{\appendixname}{End Matter}
\section{}

The utility of this composite planar extensional viscosity function is directly supported by applying it to experimental data for polymer melts and solutions. We first consider published data for the steady shear viscosity and steady planar extensional viscosity measured for Low Density Polyethylene (LDPE) at 170 $^\circ$C reported by Zatloukal \cite{zatloukal_measurements_2016}. The steady planar extensional viscosity was measured using a capillary rheometer fitted with a patented rectangular slit die. In the absence of steady $N_1$ data, we use the modified Laun's Rule \cite{das_launs_2024} to obtain an estimate for steady $N_1$ using the linear viscoelastic Small Amplitude Oscillatory Shear (SAOS) data also measured by Zatloukal \cite{zatloukal_measurements_2016}: 

\begin{equation}
    N_1 (\dot{\gamma})|_{\dot{\gamma} = \omega} \cong 2 G'(\omega) \left[ 1 + \left( \frac{G'(\omega)}{G''(\omega)} \right)^2 \right]^\ell
\end{equation}

In the empirical correlation first proposed by Laun \cite{laun_prediction_1986}, also for LDPE data, the exponent $\ell = 0.7$, but Das \textit{et al.} \cite{das_launs_2024} obtained better agreement with experimental data using $\ell = 2.23$, which was calculated by developing an extended theoretical framework that combines linear viscoelastic data, a fit to a fractional Maxwell liquid model, and nonlinear step strain data. Very good agreement between the prediction of the composite material function ($\textcolor{purple}{\text{\FiveStarOpen}}$) and actual measured steady planar extensional viscosity data ($\blacklozenge$) is achieved, as shown in Fig. 4. 

\begin{figure} [h]
    \centering
    \includegraphics[width=0.4\textwidth]{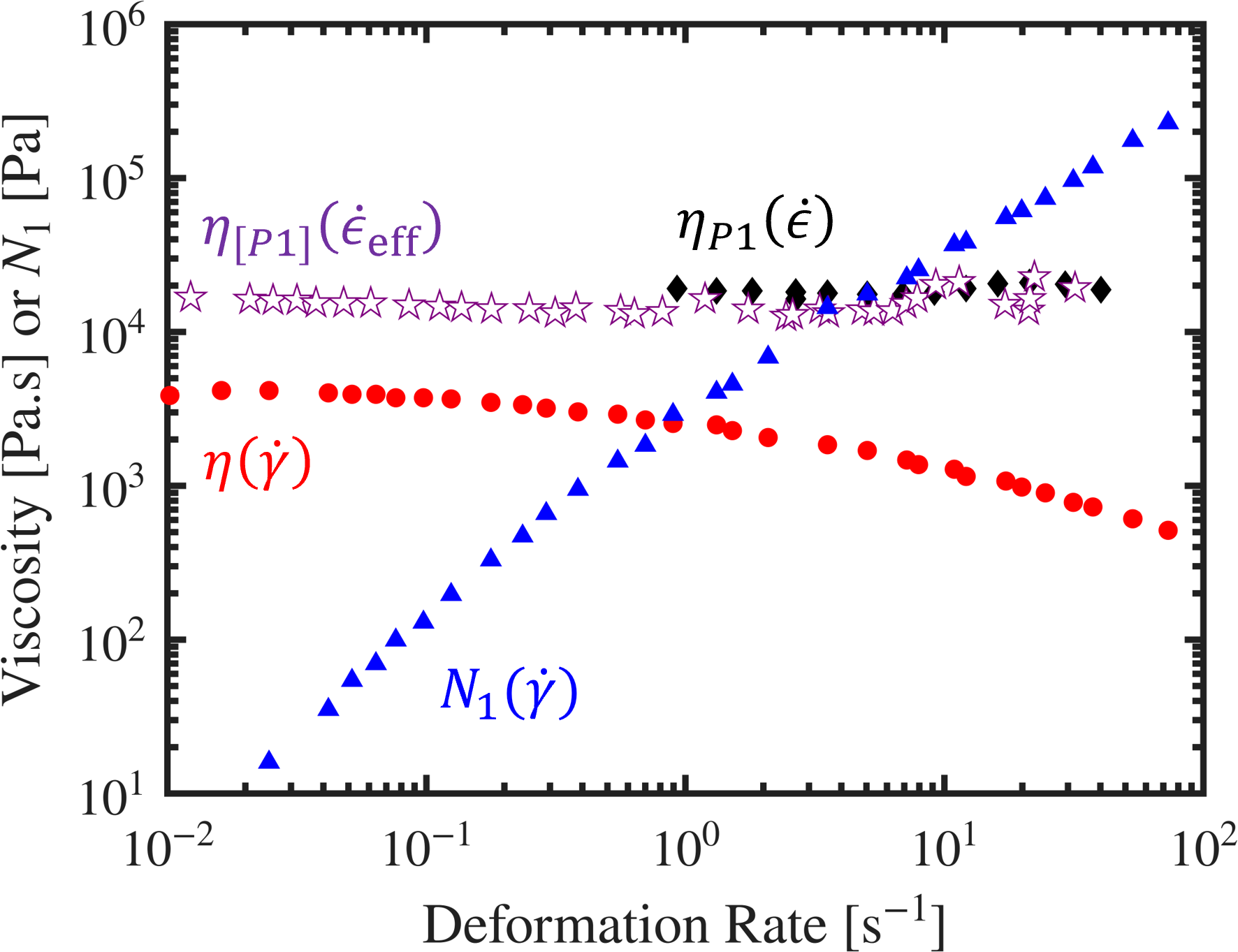}
    \caption{Experimental steady shear $\eta (\dot{\gamma})$ ($\textcolor{red}{\bullet}$) and steady planar extensional viscosity $\eta_{P1} (\dot{\epsilon})$ ($\blacklozenge$) data for an LDPE melt at 170 $^\circ$C, measured by Zatloukal \cite{zatloukal_measurements_2016}. Steady shear $N_1 (\dot{\gamma})$ data ($\textcolor{blue}{\blacktriangle}$) are predicted using an extended form of Laun's Rule with $\ell = 2.23$ \cite{das_launs_2024}. The star symbols ($\textcolor{purple}{\text{\FiveStarOpen}}$) are predictions of the new composite material function $\eta_{[P1]}(\dot{\epsilon}_\mathrm{eff})$ from Eq. \ref{etap1}.}
\end{figure}

More generally, due to the experimental difficulties in generating a steady planar extensional flow, published literature data for $\eta_{P1}$ is sparse. Semidilute entangled shear-thinning viscoelastic materials are widely used in industry as adhesives, coolants and lubricants. However, they are not viscous enough to characterize in extension using rotating clamp devices \cite{sentmanat_miniature_2004}, and are also too viscous to use in microfluidic instrumentation such as OSCER \cite{haward_optimized_2012,haward_extensional_2023} due to the large pressure drops ensuing from small device dimensions and high shear viscosity. 

We perform steady shear experiments on a 2 wt\% polyisobutylene (PIB) fluid in a paraffin oil (solvent viscosity $\eta_s = 18$ mPa.s) at 20$^\circ$C in a strain-controlled rheometer (TA Instruments ARES-G2) using a 25 mm 0.1 rad cone-and-plate fixture. %with a truncation gap of 0.045 mm 
This shear-thinning elastic polymer solution was studied in detail by More \textit{et al.} \cite{more_elasto-inertial_2024} and is in the semidilute regime, with an overlap concentration $c^* = 0.3$ wt\%. The shear data shown in Fig. 5 is truncated at the critical shear rate $\dot{\gamma}_c$ beyond which elasto-inertial instabilities onset and the flow is no longer viscometric. For illustrative purposes, we show that a single mode Rolie-Poly model is able to describe the measured steady shear viscosity ($\textcolor{red}{\bullet}$) and $N_1$ ($\textcolor{blue}{\blacktriangle}$) data. The average molecular weight of the polydisperse dissolved PIB polymer is $M \approx 10^6$ g/mol. The molecular weight between entanglements of a PIB melt $M_{e,0} \approx 7000$ g/mol \cite{unidad_consequences_2015}, the mass fraction of polymer $\phi = 2$ wt \%, and the solvent quality was determined by More \textit{et al.} \cite{more_elasto-inertial_2024} from viscoelastic measurements to be $\nu = 0.593$. The material is a `barely entangled' semidilute solution and the modified entanglement number in the solution \cite{du_capillarity-driven_2025} is $Z_{sol} = (M/M_{e,0}) \phi^{1/(3\nu-1)} \simeq 0.7$. Taking the first order truncation of the perturbation expansion by Likhtman and McLeish \cite{likhtman_quantitative_2002}, $\tau_d / \tau_R \sim 3Z = 2.1$. This is close to the best fit value $\tau_d / \tau_R = 1.3$, and rationalizes the fit.

\begin{figure} [h]
    \centering
    \includegraphics[width=0.4\textwidth]{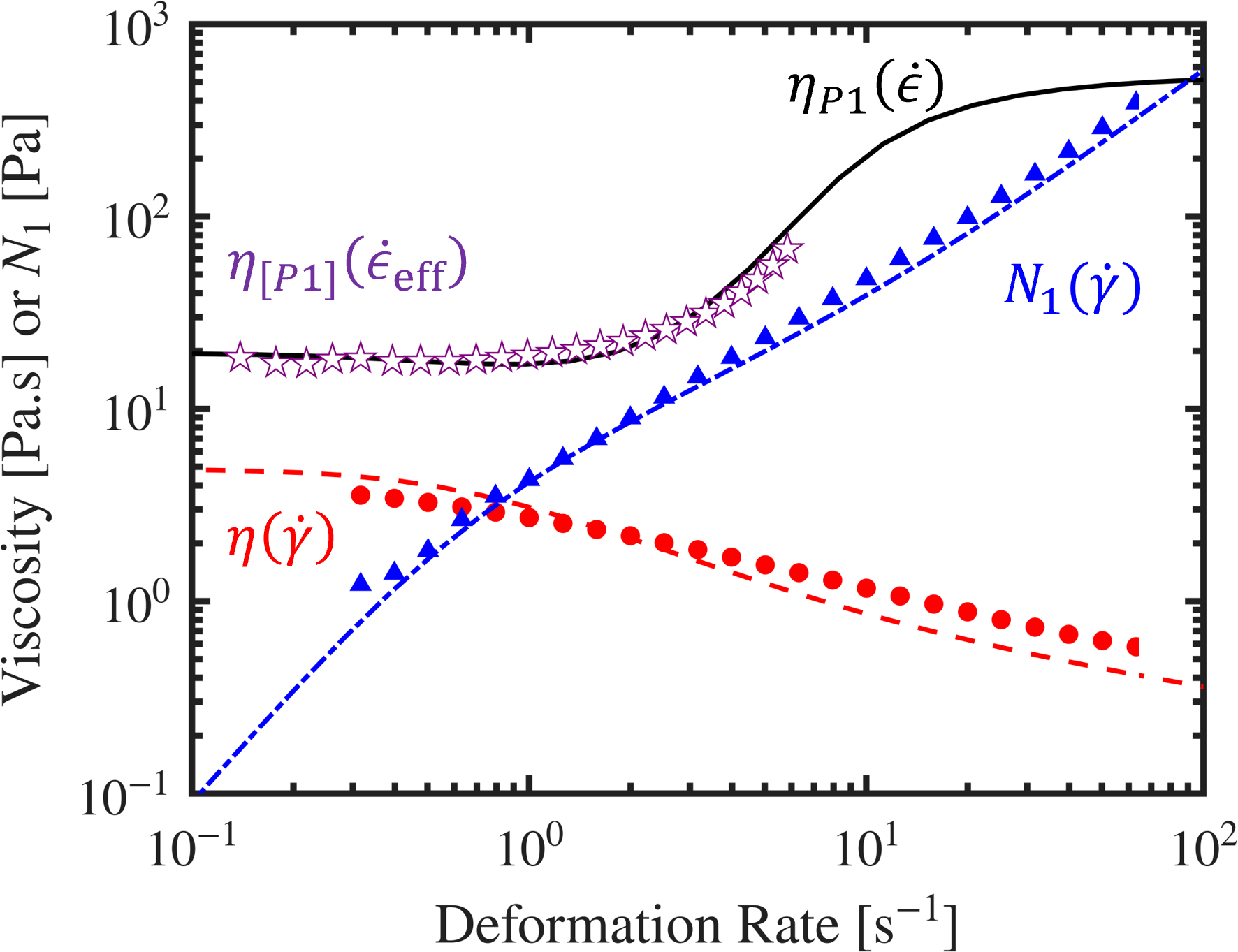}
    \caption{Experimental steady shear $\eta (\dot{\gamma})$ ($\textcolor{red}{\bullet}$) and $N_1 (\dot{\gamma})$ ($\textcolor{blue}{\blacktriangle}$) data for a 2 wt\% PIB solution. The data is fitted to a one mode Rolie-Poly model ($G = 5.18$ Pa, $\tau_d = 0.94$ s, $\tau_R = 0.75$ s, $\beta = 1.28$, $\delta = 0.305$, $L_{max} = 20$). The symbols ($\textcolor{purple}{\text{\FiveStarOpen}}$) are the new composite material function $\eta_{[P1]} (\dot{\epsilon}_\mathrm{eff})$ and the Rolie-Poly model predictions for $\eta (\dot{\gamma})$ ($\textcolor{red}{---}$), $N_1(\dot{\gamma})$ ($\textcolor{blue}{\cdot\!-\!\cdot}$) and $\eta_{P1} (\dot{\epsilon})$ (---) are shown as lines.}
\end{figure}

The Rolie-Poly model predicts that the steady planar extensional viscosity (---) starts at $4\eta_0$ at low extension rates, and begins to increase at deformation rates $\dot{\epsilon} \sim 2 - 5$ s$^{-1}$ due to the onset of chain stretching. We have shown in our theoretical analysis (Eq. \ref{eff}) that the ratio of the effective extension rate to the imposed shear rate decreases monotonically as the stresses in the fluid grow. As we show in Fig. 5 by the stars, the new composite material function $\eta_{[P1]}$, computed from measurements of the steady shear viscosity and $N_1$ data, matches very well with the predicted $\eta_{P1}$ curve when plotted as a function of $\dot{\epsilon}_\mathrm{eff}$. The Trouton ratio $Tr = \eta_{P1}(\dot{\epsilon}) / \eta (\dot{\gamma})$ increases from $Tr = 4$ at low extension rates to $Tr \simeq 50$ at $\dot{\epsilon}_\mathrm{eff} \simeq 7$ s$^{-1}$ (corresponding to the maximum shear rate $\dot{\gamma} \simeq 80$ s$^{-1}$).

\end{document}